\documentclass[a4paper]{article}
\usepackage[latin1]{inputenc}
\usepackage[dvips]{graphicx,epsfig,color}
\usepackage{wrapfig,rotating}
\usepackage{amssymb,amsmath,array}

\begin{document}
\title{\textbf{Radiation Damage of the ILC Positron Source Target\thanks{Talk presented at the Linear Collider Workshop 2007 and the International Linear Collider meeting 2007, DESY, Hamburg, 30 May -- 3 June 2007 \cite{url}.}}}
\author{Andriy Ushakov\thanks{Work supported by the Commission of the European Communities under the 6th Framework Programme ``Structuring the European Research Area'', contract number RIDS-011899.} and Sabine Riemann
\vspace{.3cm}\\
DESY \\
Platanenallee 6, D-15738 Zeuthen - Germany
}
\date{}

\maketitle
\vspace{-6cm}
\hfill DESY 07-204\\[-6mm]
\vspace{6cm}

\begin{abstract}
The radiation damage of the positron source target for the International Linear Collider (ILC) has been studied.
The displacement damage in target material due to multi-MeV photons has been calculated by combining FLUKA simulations for secondary particle production, SPECTER data for neutron displacement cross-sections and the Lindhard model for estimations of displacement damage by ions. The radiation damage of a stationary Ti6Al4V target in units of displacements per atom (dpa) has been estimated for photons from an undulator with strength  0.92 and  period 1.15 cm. The calculated damage is 7 dpa. Approximately 12.5\% of displacement damage result from neutrons.
\end{abstract}

\section{Introduction}

At the International Linear Collider (ILC) positrons will be generated by  multi-MeV photons produced in the helical undulator striking a Ti-alloy target of 0.4 radiation length thickness.  The main ILC positron source parameters (see \cite{RDR}) are shown in Table \ref{Tab:PSparameters}.

\begin{table}[h]
\begin{center}
\begin{tabular}{|l|c|}
\hline
e$^-$ drive beam energy         & 150 GeV           \\ \hline
Undulator K-value               & 0.92 \\ \hline
Undulator period                &1.15 cm \\ \hline
Undulator-target distance       & 500 m          \\ \hline
e$^+$ per bunch at IP           & $2 \cdot 10^{10}$ \\ \hline
Bunches per pulse               & 2625 \\ \hline
Pulse repetition rate           & 5 Hz          \\ \hline
\end{tabular}
\end{center}
\caption{ILC positron source parameters \cite{RDR}.}
\label{Tab:PSparameters}
\end{table}

The photons passing through the target generate not only electron-positron-pairs that are captured and accelerated downstream the target but also many other secondary particles (electrons, protons, neutrons, ions and etc.). 
In the target material the secondary particles collide with lattice atoms called ``Primary Knock-on Atoms'' (PKAs). If the energy transferred in collisions is higher than a displacement threshold energy, $E_d$, the atoms leave their  site and start moving. During its movement, a PKA is slowed-down by interactions with other atoms, which are called ``Secondary Knock-on Atoms'' (SKAs). If the PKA energy is high enough (few tens keV for metals~\cite{Stoller}), PKAs and SKAs will produce a displacement cascade or even many cascades. Most of the created vacancies and self-interstitial atoms annihilate each other. At the end of this recombination phase, only some residual point defects remain. 

The goal of this article is to estimate the radiation damage of the positron source target by multi-MeV photons generated in a helical undulator.

\section{Damage modeling}

The photons are generated by the radiation of
relativistic electrons 
passing through the periodic, helical magnetic field of the undulator. The undulator is installed part way along main electron linac, in a chicane arrangement, where the electron energy has reached 150~GeV. Positrons are produced from multi-MeV photons by means of pair production in the field of a nucleus in a conversion target. 
When a nucleus is excited by photons at energies of giant resonances, one or more nucleons may  knocked out. 
These secondary particles collide with lattice atoms and  induce damage of the target. 

\begin{figure}[ht]
\begin{center}
\includegraphics[width=5.9cm,keepaspectratio=true]{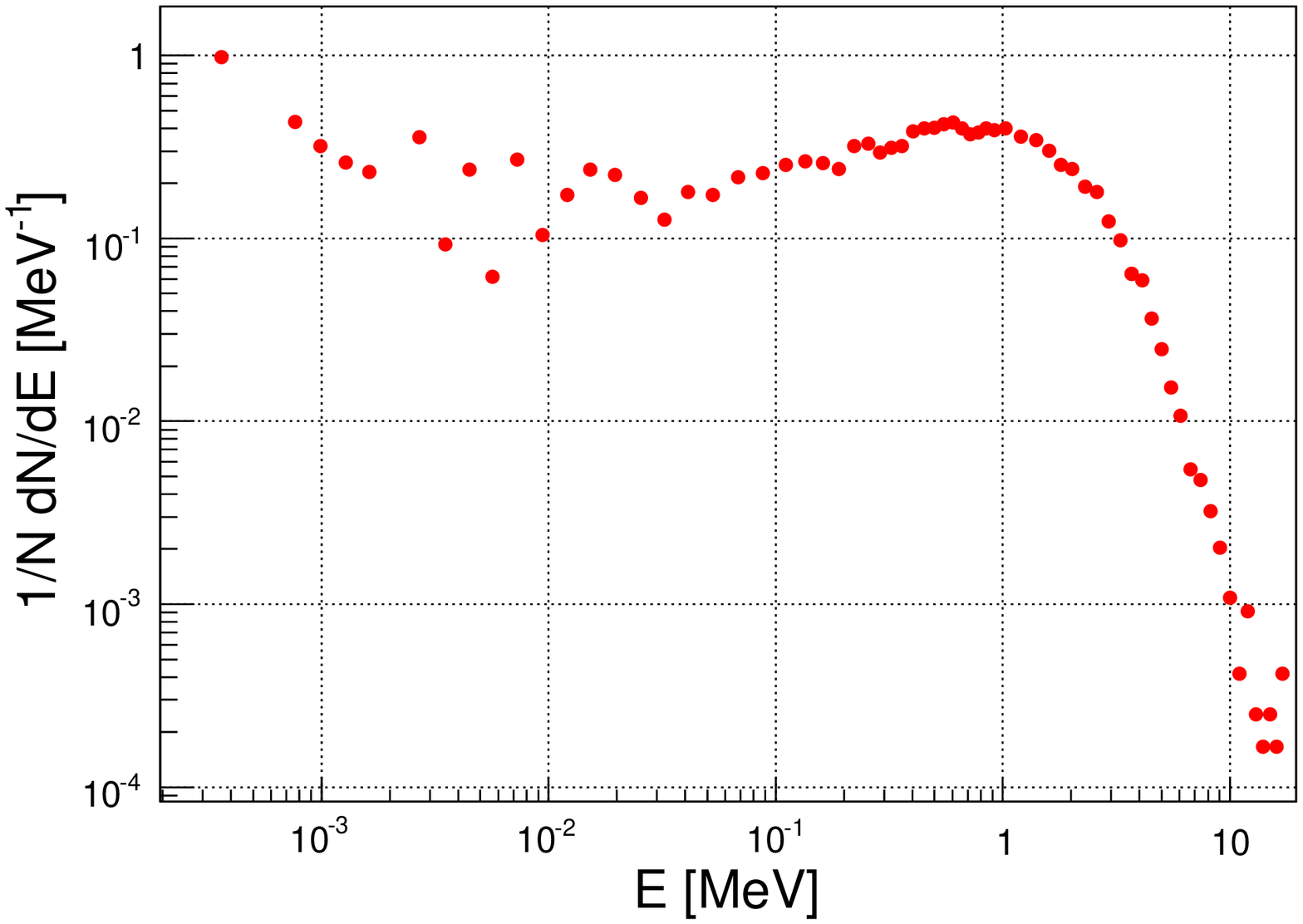}$~~~~$
\includegraphics[width=5.55cm,keepaspectratio=true]{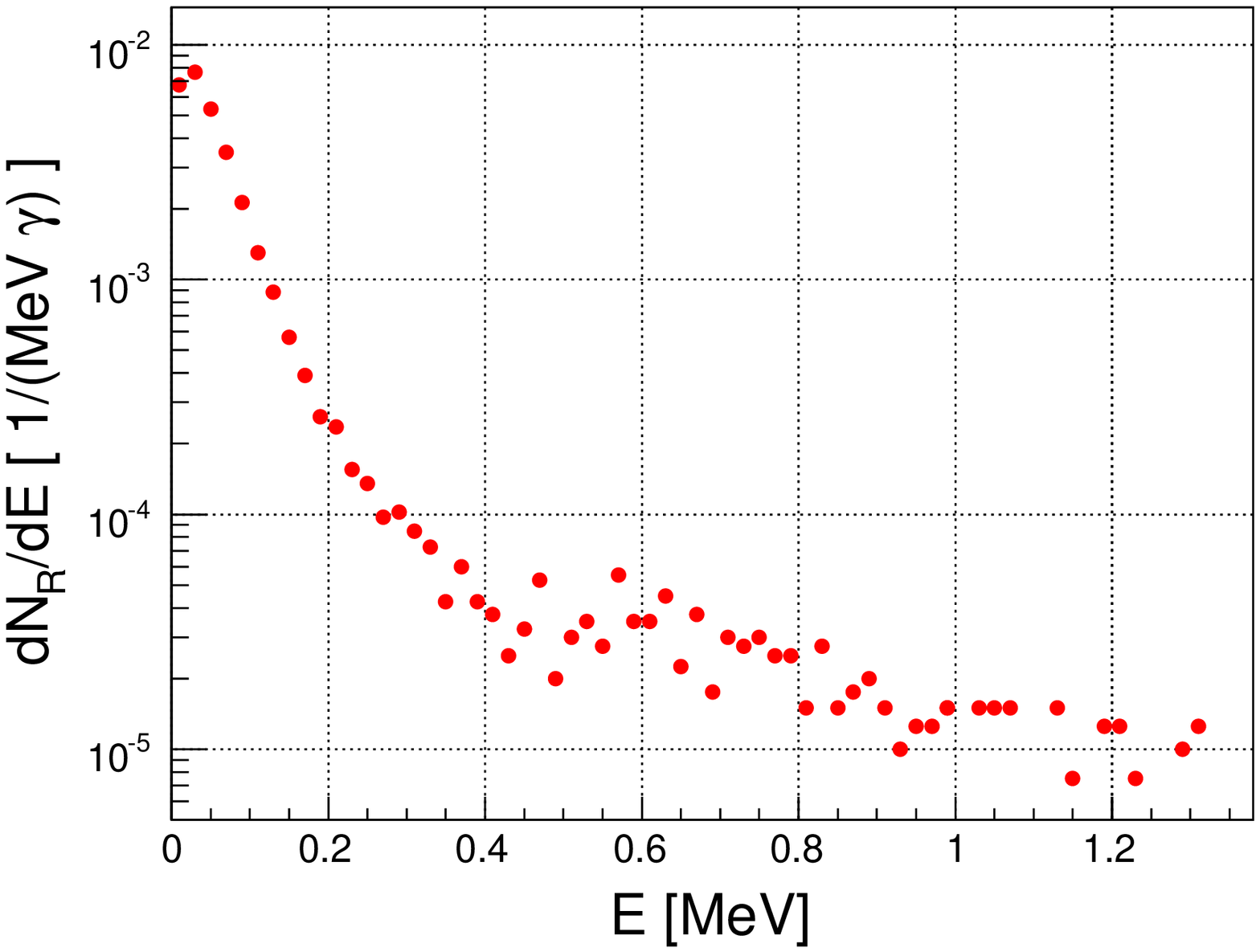}
\caption{Energy distribution of neutrons (left) and recoil nuclei (right) in the Ti-alloy target.}
\label{Fig:NEDF}
\end{center}
\end{figure}

The generation of secondary particles by a photon beam has been calculated using FLUKA~\cite{Fasso}.
A major advantage of FLUKA is its capability to calculate electromagnetic showers in matter as well as 
neutron fluxes and atomic recoil effects. The Monte-Carlo code is well suited to describe 
residual nuclei in electromagnetic and hadronic cascade processes. The FLUKA output data for secondary photons, neutrons, protons, deutrons, tritions, $^3$He, alpha particles and recoil nuclei from the nuclear reactions have been used for radiation damage calculations. The FLUKA output includes the positions and momenta 
of secondary particles. Fig. \ref{Fig:NEDF} shows neutron energy distribution 
and energy spectrum of recoil nuclei in the Ti-alloy target generated by undulator photons (see Table \ref{Tab:PSparameters}).

\begin{figure}[h]
\begin{center}
 \includegraphics[width=8cm,keepaspectratio=true]{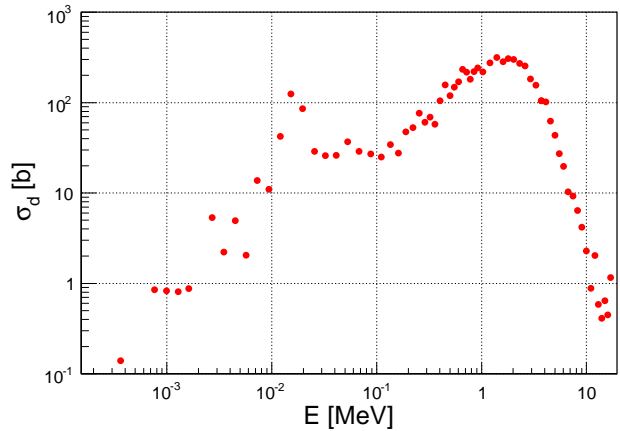}
\end{center}
\caption{Neutron damage cross-section, $\sigma_d$, induced by a photon beam generated in a helical undulator with $K = 0.92$ and $\lambda = 11.5$ mm.}
\label{Fig:sigma-d}
\end{figure}

Following the approach used in \cite{Wirth} the calculations of diplacements in the target material have been performed seperately for  damage induced by neutrons, ions and recoil nuclei.
For estimations of displacements induced by neutrons the energy dependent displacement cross-sections from SPECTER code have been used \cite{SPECTER}. 
Fig.~\ref{Fig:sigma-d} shows the neutron damage cross-section  $\sigma_d$ 
created by the photon beam generated in a helical undulator with $K = 0.92$ and $\lambda = 11.5$ mm.

The number of atoms  displaced by neutrons ($dpa_n$) is the product of  $\sigma_d$ and total neutron fluence $\phi_n$:
$
 dpa_n = \sigma_d \, \phi_n
$.

\begin{figure}[ht]
\begin{center}
\includegraphics[width=9.5cm,keepaspectratio=true]{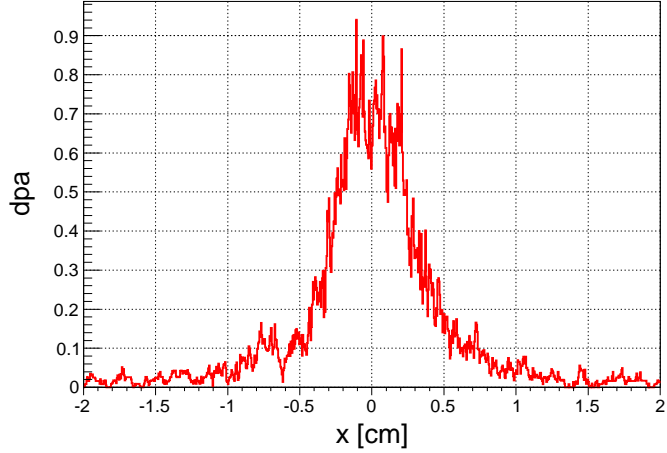}
\caption{Transverse space distribution of neutron induced dpa in a  Ti6Al4V target after 5000 hours irradiation.}
\label{Fig:dpaN-vs-x}
\end{center}
\end{figure}

In order to estimate the target damage it is necessary to know the power of the photon beam incident on the target. 
The required photon beam power has been derived from the  positron production in the Ti6Al4V target (90\% of Ti, 6\% of Al and 4\% of V) with a thickness of 0.4 radiation length and  taking into account  positron beam losses after the target.
The positron yield defined as ratio of number of positrons emerging from the target and number of photons hitting the target is $2.25 \cdot 10^{-2}$. Almost 28\% of the positrons are captured, accelerated and match
the acceptance of the dumping ring. 
Hence, for the nominal ILC positron source parameters (Table \ref{Tab:PSparameters})  $7 \cdot 10^{16}$ photons per second are required corresponding to 117 kW average photon beam power. 
Fig.~\ref{Fig:dpaN-vs-x} shows the transverse dpa distribution induced by neutrons after 5000 hours of source operation averaged over the full target thickness.

For the calculation of displacements induced by ions and recoil nuclei the Linhard model \cite{Robinson} has been  used. 
The dpa produced by ions, $dpa_i$, has been determined as 
\[
dpa_i = \sum\limits_j \frac{{\hat E_j}}{L}\frac{\phi_j }{{N_a }},
\]
where $\phi_j$ and $\hat E_j$ are fluence and damage energy of j-type projectiles (protons, deutrons, tritions, $^3$He, alpha particles, heavy particles and recoil nuclei), and $N_a$ is the number of target atoms per cm$^2$.  $L$ is the cascade multiplication threshold,

\[
L = E_b + E_c + E_d,
\]
where $E_b$ is the energy binding an atom to its lattice site, 
$E_c$ is the energy of the slow projectile to be captured  by a vacant lattice site,
$E_d$ is the energy for displacing an atom permanently from its lattice site. The atom could be displaced if $\hat E \geq L$. In our estimations $L = 2 E_d / 0.8 $ has been used as recommended by \cite{Robinson}.
$E_d=40~$eV  for Ti, V and $E_d=27~$eV  for Al \cite{SPECTER}. The damage energy has been approximated as (see \cite{Robinson})
\[
\hat E = \frac{E}{{1 + k_L (\varepsilon  + 0.40244\varepsilon ^{3/4}  + 3.4008\varepsilon ^{1/6} )}},
\]
where $E$ is the energy of projectile particle, and $\varepsilon$ is a dimensionless energy,
\[
\varepsilon  = \frac{E}{{E_L }},\quad E_L  = \frac{{Z_1 Z_2 e^2 }}{{a_{12} }}\frac{{1 + A}}{A},
\]
\[
k_L  = \frac{{32}}{{3\pi }}\sqrt {\frac{{m_e }}{{M_2 }}} \frac{{(1 + A)^{3/2} Z_1^{2/3} Z_2^{2/3} }}{{(Z_1^{2/3}  + Z_2^{2/3} )^{3/4} }}. 
\]
$Z_1$ and $M_1$ are the atomic number and mass of the projectile, $Z_2$ and $M_2$ are the atomic number and mass of the target atom, $A = M_2/M_1$, $e$ is the charge of the electron, $a_{12}$ is a Thomas-Fermi screening length of the atom \cite{Lindhard}
\[
a_{12}  = \left( {\frac{{9\pi }}{{128}}} \right)^{1/3} \frac{{a_H }}{{(Z_1^{2/3}  + Z_2^{2/3} )^{1/2} }},
\]
where $a_H$ is the Bohr radius.

\begin{figure}[ht]
\begin{center}
\includegraphics[width=6.5cm,keepaspectratio=true]{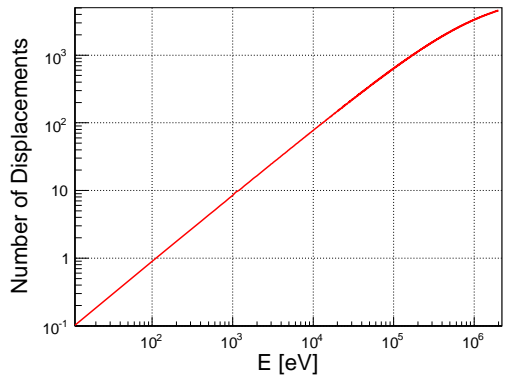}
\caption{Number displaced atoms by $^{48}$Ti recoils in Ti6Al4V target as function of $^{48}$Ti energy.}
\label{Fig:Ndisplacements-by-Ti}
\end{center}
\end{figure}

\begin{figure}[h]
\begin{center}
\includegraphics[width=6cm,keepaspectratio=true]{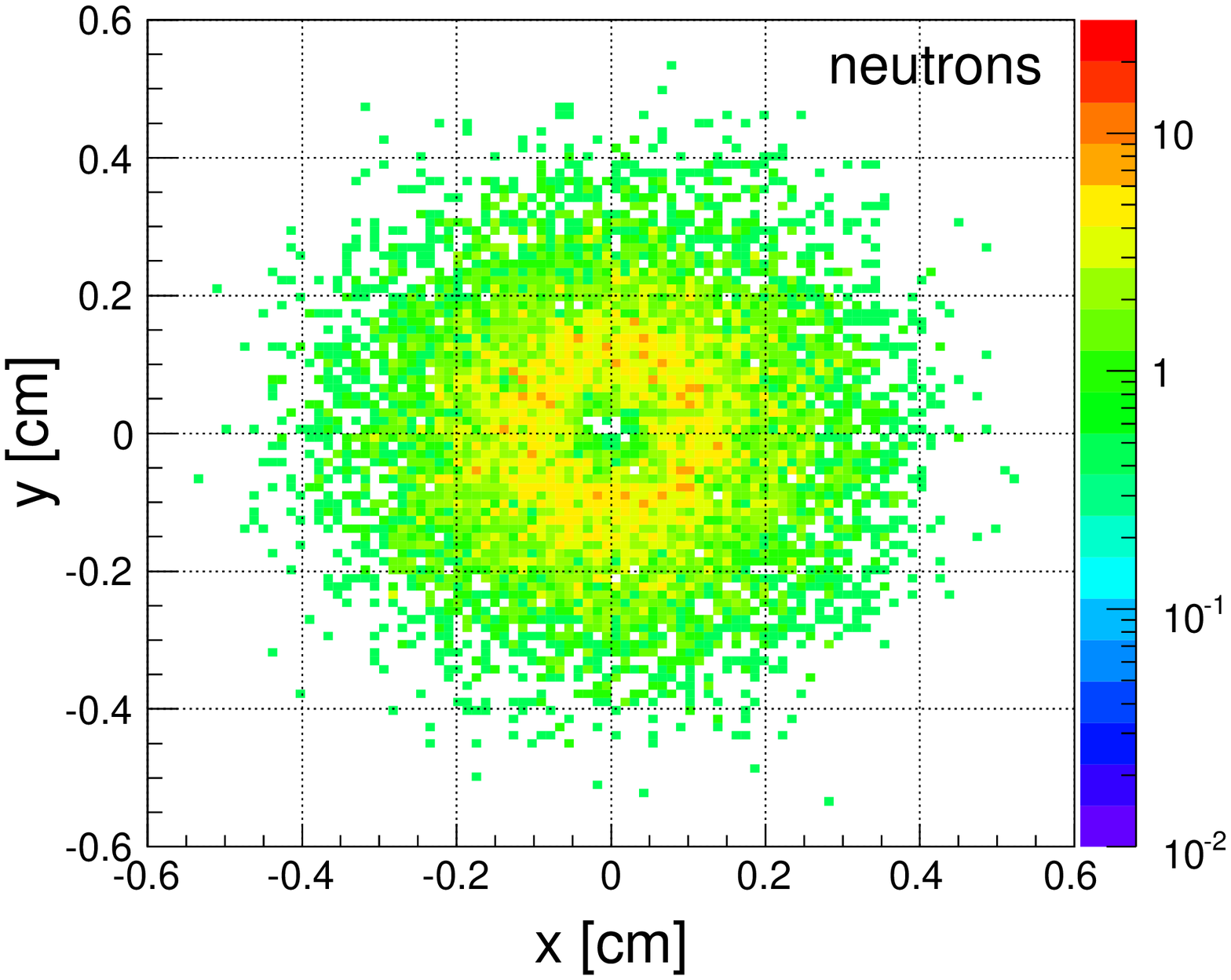}
\includegraphics[width=6cm,keepaspectratio=true]{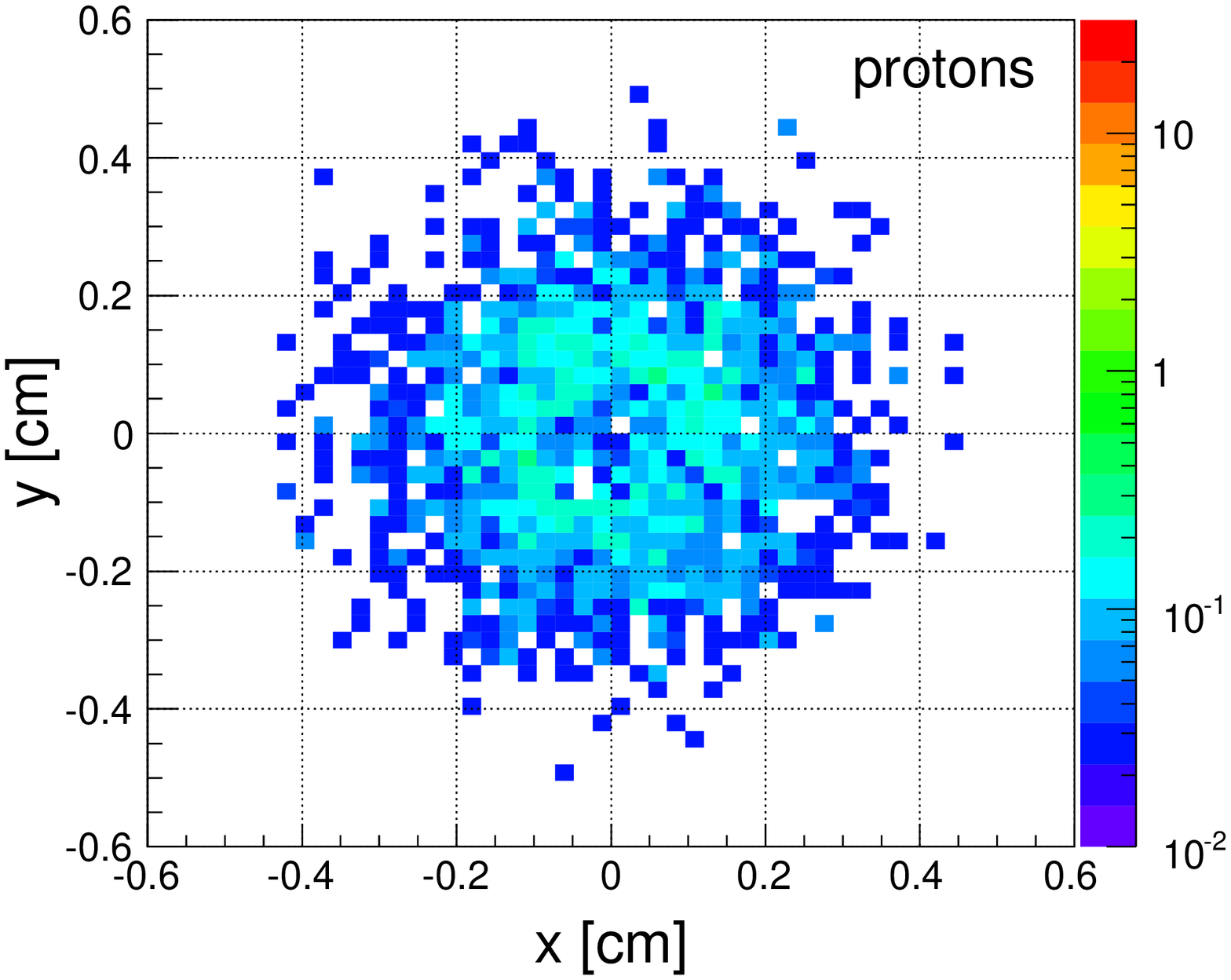}
\includegraphics[width=6cm,keepaspectratio=true]{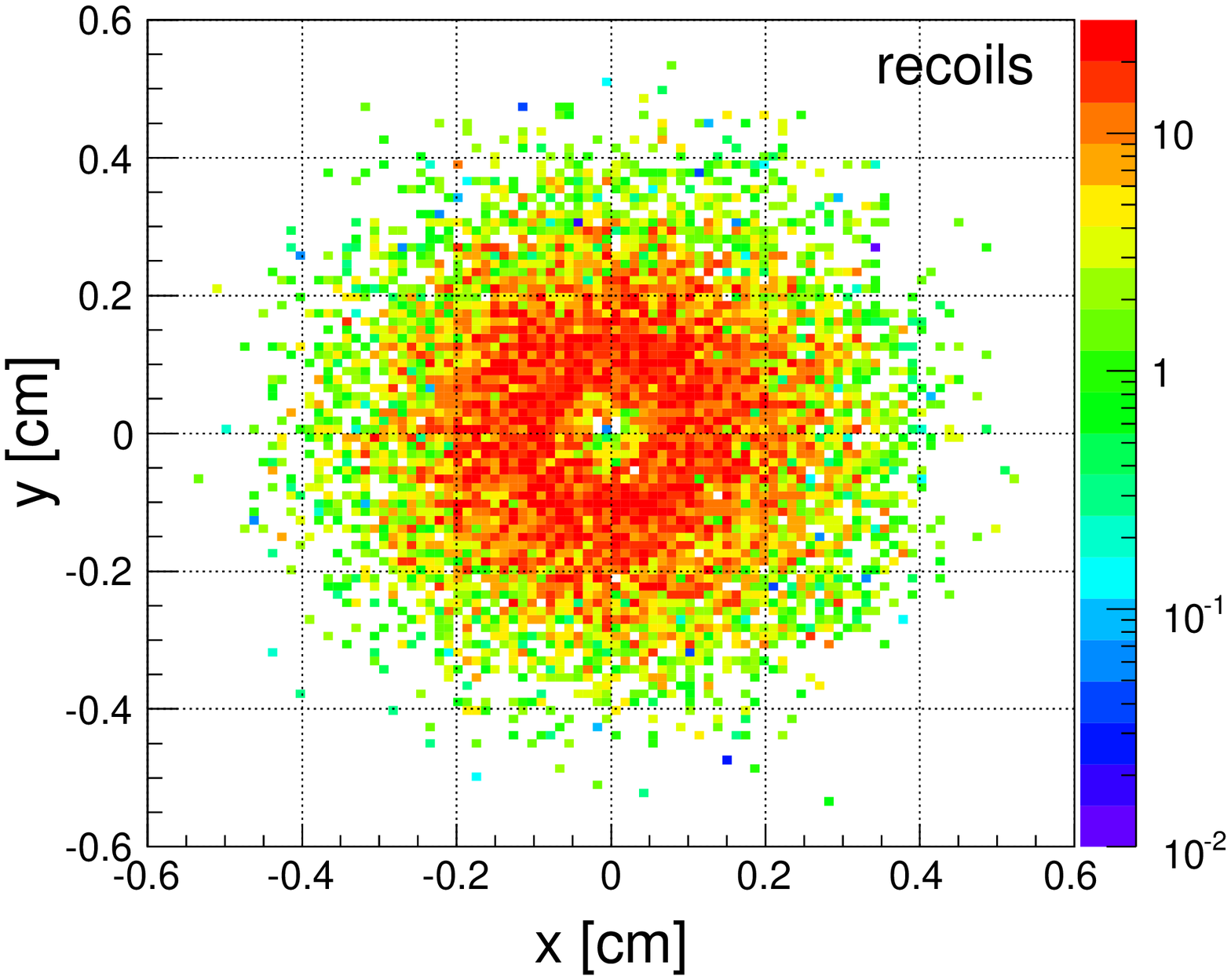}
\includegraphics[width=6cm,keepaspectratio=true]{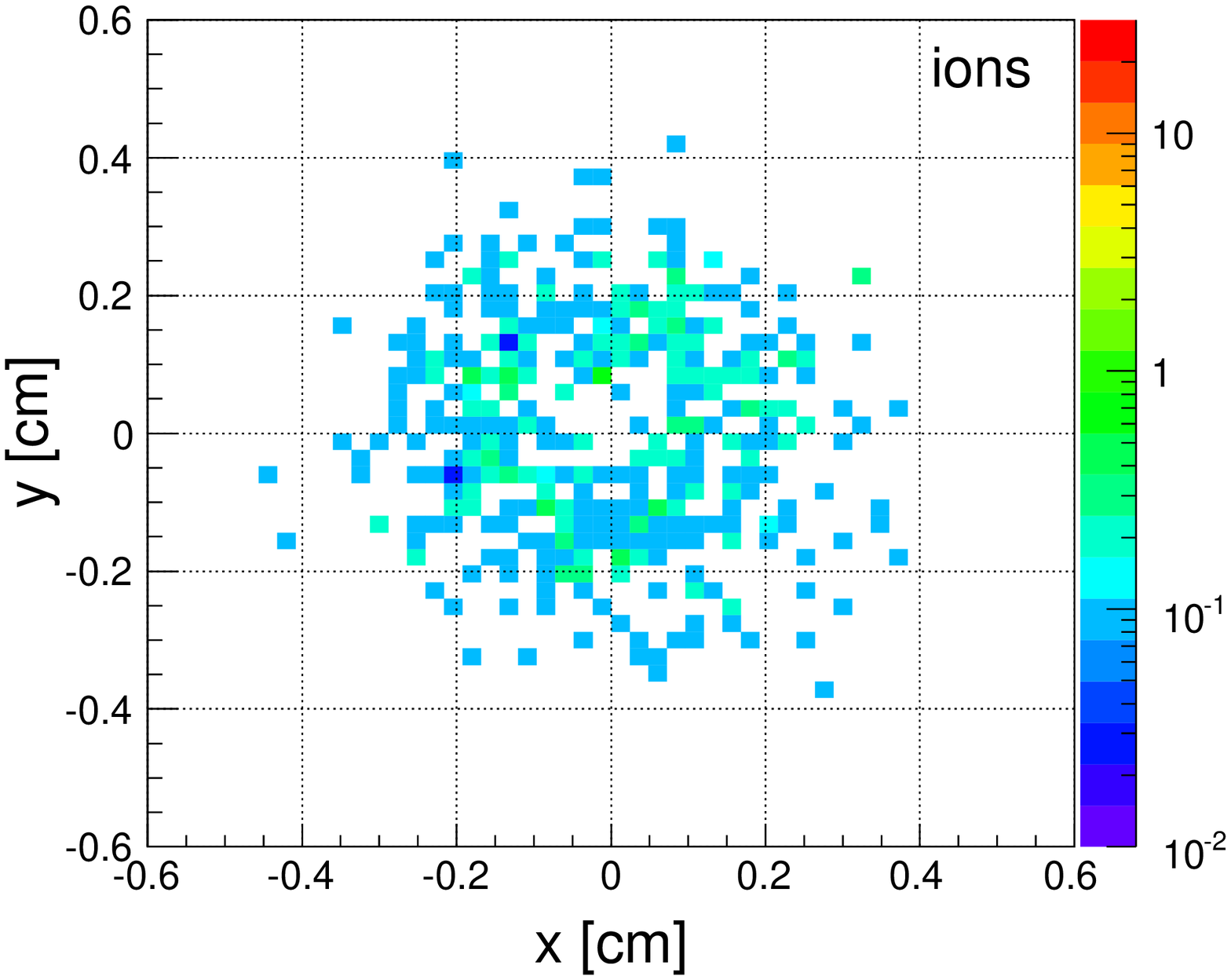}
\caption{Average dpa induced by neutrons, protons, ions and recoil atoms (without taking into account  secondary particle transport).}
\label{Fig:dpa-vs-xy}
\end{center}
\end{figure}

The number of displaced atoms in the  Ti6Al4V target induced by $^{48}$Ti recoils 
is shown in Fig. \ref{Fig:Ndisplacements-by-Ti} as function of $^{48}$Ti energy. Figure \ref{Fig:dpa-vs-xy} shows  dpa averaged over target thickness induced by neutrons, protons, ions (deutrons, tritions, $^3$He, alpha particles, heavy particles) and all recoil atoms after 5000 hours of source operation. These results are quantitative only, because the ``damage area'' used for the estimations has been defined 
by the size of the photon beam. This area has been frozen after the first inelastic photon-nucleus interaction and does not include the real displacement cascade development. The reason for such essential simplification was the inability of FLUKA to transport 
recoil atoms. 
Hence, the results shown in Figure \ref{Fig:dpa-vs-xy} could be used only for relative comparisons. 
 
Displacements induced by neutrons are one order of magnitude less then dpa by  recoil atoms;  damage by protons and ions (deutrons, tritions, $^3$He, alpha particles and heavy particles) is smaller by two order of magnitude. 
Taking into account the transport of recoil atoms will increase the damage area essentially or decrease the fluence of recoil atoms and results in a reduced dpa value. 
That is demonstrated in Fig. \ref{Fig:dpa-vs-xy} and Fig. \ref{Fig:dpaN-vs-x}:
Considering only neutrons  and neutron transport 
the  maximal dpa value is reduced from 8.7 to 0.9 
(scaling factor is approximately 10). The estimated total dpa value induced by all secondary particles  and their transport is 7 dpa after applying the scaling factor.

\section{Summary}

The radiation damage of stationary  target for an  ILC positron source has been estimated using a simplified model.
The Ti6Al4V target was hitted by  117 kW photon  generated in a helical undulator with  strength 0.92 and  period 1.15 cm. 
The calculated damage is about 7~dpa after 5000 hours of positron source operation. The main contribution to dpa is induced by recoil atoms, only approximately 12.5\% of displacement damage result from neutrons. 

\begin{footnotesize}

\end{footnotesize}

\end{document}